\documentclass[letterpaper,twocolumn,10pt]{article}
\usepackage[hyphens]{url}
\usepackage{soul}
\usepackage{balance}
\usepackage{usenix,epsfig,hyperref,caption,subcaption}
\hypersetup{
    bookmarksnumbered=true,
    unicode=false,          
    pdftoolbar=true,        
    pdfmenubar=true,        
    pdffitwindow=false,     
    pdfstartview={FitH},    
    pdftitle={Censorship is Futile},    
    pdfnewwindow=true,      
    colorlinks=false,       
    linkcolor=red,          
    citecolor=green,        
    filecolor=magenta,      
    urlcolor=cyan           
}
\begin{document}

\date{}

\title{\Large \bf \st{Resistance} Censorship is Futile}

\author{
{\rm Zubair Nabi}\\
IBM Research -- Ireland\\
zubairn@ie.ibm.com
}

\maketitle


\thispagestyle{empty}

\begin{abstract}
The Internet has become the new battle ground between authoritarian regimes and
ordinary individuals who want unimpeded access to information. The immense
popularity of online activism and citizen journalism enabled by social media has
instigated state level players to partially or completely block access to the
Internet. In return, individuals and organizations have been employing various
anti-censorship tools to circumvent these restrictions. In this paper, we claim
that \emph{censorship is futile} as not only has it been ineffective in
restricting access, it has also had the side-effect of popularising blocked
content. Using data from Alexa Web Rankings, Google Trends, and YouTube
Statistics, we quantify the ineffectiveness of state level censorship in
Pakistan and Turkey and highlight the emergence of the \emph{Streisand Effect}.
We hope that our findings will, a) prove to governments and other players the
futility of their actions, and b) aid citizens around the world in using legal
measures to counteract censorship by showing its ineffectiveness.
\end{abstract}

\section{Introduction}
The popularity of the Internet is a double-edged sword: it opens up a world of
information and promotes free speech but at the same time, this popularity
incites authoritarian regimes and other actors to repress access to it. As a
result, at present more than 60 countries around the world censor the Internet
in one form or another~\cite{Burnett:2013:MSI}. These forms include all-out
blocking~\cite{Dainotti:2011:ACI}, partial blocking at different
levels~\cite{Nabi:2013:TAO,Verkamp:2012:IMO,Aryan:2013:ICI}, performance
degradation/throttling~\cite{Anderson:2013:DTI}, and content
manipulation~\cite{Bamman:2012:CAD,Verkamp:2013:FIO}. To stem this suppression
tide, researchers and activists are actively developing an arsenal of various
anti-censorship
tools~\cite{Houmansadr:2011:CCI,Wustrow:2011:TAN,Dingledine:2004:TSO,Feamster:2002:ICW,Burnett:2010:CAC,Clarke:2002:PFE,Invernizzi:2013:MBS}.
Cognizant of this, censors are trying to stay ahead of the
curve~\cite{Winter:2012:HTG,Schuchard:2012:RAD,Geddes:2013:CYA}.
This has led to a cat and mouse game between censors and anti-censorship
practitioners, with no end in sight, similar to the struggle between
encryption-decryption~\cite{Barto:2013:EAD}.

Fortunately, this tug of war has not dampened the desire of users around the
world to gain unimpeded access to the Internet. On the one hand, they are
readily using various methods to circumvent restrictions~\cite{Nabi:2013:TAO}
and on the other they are helping others to do the same through ingenious
mechanisms, such as putting up addresses of open DNS servers as wall
graffiti~\cite{Zmijewski:2014:TIC}. In a similar vein, in spite of restrictions
and the risk associated with bypassing them, the usage of social media and blogs
has been escalating~\cite{Caren:2014:PCH}. At the other end of the spectrum, all
of this online activism has induced some governments to pass ``Internet
Constitutions'' to enshrine freedom of expression, net neutrality, and online
privacy~\cite{Mari:2014:BPG}.

Essentially, at the very core, the goal of a censor is to restrict access to
content that is deemed detrimental to a certain vested interest. This blackout
often backfires and causes the content to go viral following the Streisand
Effect~\cite{Greenberg:2007:TSE}. The Streisand Effect can be defined as the
inadvertent popularity of any material as a result of its suppression (Details
in \S\ref{subsec:streisand}). In recent years a number of instances of this
phenomenon have been recorded. These include the swell in the number of Twitter
users after the website was blocked in Turkey~\cite{Woollacott:2014:SET}, the
popularity of the online activism portal Avaaz's posters to ban South Africa's
lion bone trade~\cite{Solon:2013:ACA}, and the re-posting of the ``Station
hertzienne militaire de Pierre-sur-Haute'' entry on Wikipedia after the latter
was forced by French Intelligence to delete it~\cite{Geuss:2013:WEA}. While the
Internet is replete with examples of this phenomenon, the evidence is mostly
anecdotal and not in the context of state-level censorship.

\begin{table*}[t]
\centering
    \begin{tabular}{| l | l | p{11cm} |}
    \hline
    \textbf{Date} & \textbf{Event ID} & \textbf{Description} \\
	\hline
	February 22, 2008 & \emph{PK1} & YouTube censored in retaliation to the movie
	``Fitna''\\
	May 19, 2010 & \emph{PK2} & Facebook, YouTube, Flickr, and Wikipedia
    blocked due to ``Everybody Draw Muhammad Day''\\
    September 17, 2012 & \emph{PK3} & YouTube blocked in reaction to another
    controversial  movie ``Innocence of Muslims''\\
    \hline
    \end{tabular}
    \caption{Censorship Events in Pakistan}
    \label{tab:pkevents}
    \vspace{-5pt}
\end{table*}

\begin{table*}[t]
\centering
    \begin{tabular}{| l | l | p{11cm} |}
    \hline
    \textbf{Date} & \textbf{Event ID} & \textbf{Description} \\
	\hline
	May 08, 2008 & \emph{TR1} & YouTube blocked due to videos offensive to Mustafa
	Kemal Ataturk\\
	January 27, 2014 & \emph{TR2} & SoundCloud blocked to stop access to leaked
	audio tapes of the Turkish Prime Minister and his main political rival in a
	possible graft case \\
	March 20, 2014 & \emph{TR3} & Twitter blocked to suppress
	access to leaked recordings allegedly implicating the Turkish Prime Minister and his inner
	circle in various corruption scandals\\
    March 27, 2014 & \emph{TR4} & YouTube censored to block access to audio
    recordings of government officials discussing a possible military strike
    inside Syria\\
    \hline
    \end{tabular}
    \caption{Censorship Events in Turkey}
    \label{tab:trevents}
    \vspace{-5pt}
\end{table*}

To remedy this, we present, to the best of our knowledge, the first formal study
of the Streisand Effect in state-level censorship. With Pakistan and Turkey as a
case study, using data from Google Trends, YouTube Video Statistics, and Alexa
Web Rankings, we show that censorship has, a) not affected the ranking of
websites such as YouTube, and b) it has had the side-effect of causing
restricted content to go viral. The goal of this paper is two-fold: to dissect
the Streisand Effect in the context of state-level censorship and also, possibly
more importantly, to aid citizens around the world in using legal measures to
counteract censorship. For instance, for the past two years activists have been
fighting a case in a provincial High Court in Pakistan to unblock YouTube.
The ineffectiveness of the ban has even prodded the lead judge to remark at one
point: ``What the government is saying is a joke. Every child is accessing
YouTube. It's a fraud against the people of Pakistan''~\cite{Shah:2014:TTD}.
Unfortunately, most of the information about censorship circumvention and the
Streisand Effect is circumstantial and anecdotal, to which the law is blind.
Therefore, a formal study with concrete numbers can fortify the case for the
futility of Internet censorship around the world. It has already proven
effective in countries like Turkey where the ban on Twitter was lifted recently
after the constitutional court ruled it as illegal~\cite{Reuters:2014:TLT}.

The rest of the paper is organized as follows. In \S\ref{sec:background} we give
background information about the Streisand Effect and the various data sources
used in the paper. \S\ref{sec:results} presents our findings. We discuss the
implications of the findings in \S\ref{sec:discussion}. \S\ref{sec:conclusion}
concludes the paper and also discusses future directions. 

\section{Setting The Stage}\label{sec:background}

In this section, we give the reader background information about the Streisand
Effect, Alexa Web Rankings, Google Trends, and YouTube Video Statistics. In
addition, we give details of Internet censorship in our two target countries:
Pakistan and Turkey.
\subsection{Streisand Effect}\label{subsec:streisand}
The Streisand Effect is named after the singer Barbara Streisand who in 2003
unsuccessfully tried to have an aerial picture of her Malibu beach house taken
down from the website of an environmental activist~\cite{Greenberg:2007:TSE}.
She subsequently lost the \$50 million lawsuit but the ensuing media coverage
brought more than a million visitors to the said website. Since then, the
Streisand Effect has become synonymous with the unintentional virality of any
information, online or otherwise, as a consequence of any attempt to censor,
suppress, and/or conceal it. This phenomenon has manifested itself in a diverse
array of settings. Examples include unsuccessful lawsuits and
injunctions~\cite{Solon:2013:FBT,Jacbonson:2009:TCN}, personal
liberty~\cite{Cacciottolo:2012:TSE}, and even scientific
research~\cite{Heussner:2008:ITS}. Over the years, the effect has been
leveraged by individuals and organizations alike to repel censorship. For
instance, the American Library Association and others have been organizing a
Banned Books week since 1982 to raise awareness about banned and restricted
books~\cite{Doyle:2010:BBC}. Similarly, attorneys frequently advise their
clients not to pursue libel cases that can potentially raise interest in the
issue rather than subdue it~\cite{Eagle:2012:TSE}. Above all, the Streisand
Effect has been successfully used by numerous individuals around the world to
take the fight to government mandated
censorship~\cite{Economist:2008:AGC,Woollacott:2014:SET,Greenberg:2007:TSE}. It
is important to highlight that while the positive effects of the effect are an
effective tool against censorship, it is by no means a panacea. To counteract
online censorship around the world, we still require an ecosystem of activists
and tools. We discuss this in detail in \S\ref{subsec:means}.

\subsection{Pakistan}
With a population size almost half of South America and an almost comparable
number of absolute Internet users~\cite{Nabi:2013:TAO,IWS:2011:IUA}, Pakistan
has experienced Internet censorship in many forms over the years. Popular
websites such as Google, Facebook, Flickr, Wikipedia, and YouTube have borne the
brunt of restrictions in recent years. At times, the side-effects of this
censorship have affected connectivity outside the country as
well~\cite{Brown:2010:PHY}. From a technical perspective, websites are blocked
at both the DNS level as well as the HTTP level~\cite{Nabi:2013:TAO}. These
restrictions are imposed through technology from Netsweeper at the IXP
level~\cite{Citizen:2013:OPW}. In addition, citizens are also spied upon through
FinSpy~\cite{Boire:2013:FTE}. Fortunately, Pakistan also has an extremely
vibrant social activism community which has legally challenged Internet
censorship and surveillance within the
country~\cite{Shah:2014:TTD,Malik:2014:PFA}. Table~\ref{tab:pkevents} shows
three major censorship events from the country under consideration in this
paper.

\subsection{Turkey}
An Internet penetration of close to 49\% makes Turkey one of the most connected
countries in the world~\cite{eMarket:2013:YDW}. It also has the highest Twitter
penetration in the world~\cite{Baydar:2014:TIB}. In addition, 72\% of the
Internet users access newspapers and magazines online~\cite{TSI:2012:ICT}
enabling them to be politically informed. Unfortunately in the last few years,
Turkey has gone from relatively benign online censorship to ``one of the world's
most determined Internet censors''~\cite{Parkinson:2014:TEO}. In fact, the
Turkish parliament recently passed legislation that allows the authorities to
block any website while bypassing the courts~\cite{BBC:2014:TPL}. In addition,
Internet service providers are bound to keep tabs on a user's online activity
for two years for use by the authorities. Similar to Pakistan, Turkey also has a
very active online community whose activism recently prompted its top court to
deem the ban on Twitter as illegal~\cite{Reuters:2014:TLT}. Four main
censorship events from Turkey are listed in Table~\ref{tab:trevents}.

\subsection{Alexa Web Rankings}
Alexa\footnote{\url{http://www.alexa.com/}} is a web analytics company that
provides traffic analysis and rankings for top level domains. Traffic
rankings--which are updated daily--are calculated by using crowd-sourced data
from its browser toolbar. A website's rank is a function of the number of unique
visitors and page views. Alexa's web portal displays rankings via three views:
1) global, 2) country-wise, and 3) category-wise. In addition, Alexa also
exposes a \emph{Web Information Service} API that allows the user to query
historical data through Amazon Web Services.

\subsection{Google Trends}
Google Trends\footnote{\url{http://www.google.com/trends/}} is a time series
index of search term volume entered into Google in a geographic
region~\cite{Choi:2012:PTP}. The index is calculated by normalizing the total
query volume for a specific term by the total number of queries in that region.
The maximum value is 100 and indexes can be searched back till January 1, 2004.
Google Trends have found wide traction for trend prediction, from economic
indicators~\cite{Choi:2012:PTP} to disease epidemics~\cite{Ginsberg:2009:DIE}.
Similar to Alexa, Google Trends has both a web-based portal as well as a
back-end API.

\subsection{YouTube Video Statistics}
Statistics for a YouTube video can be enabled by the uploader for public
consumption. These statistics can include a time series of views, duration of
views, and shares. Instead of sharing raw numbers, the portal only exposes
generated charts of regular time series or CDFs. In addition, no external API is
available to access this information.

\begin{table}[t]
\centering
    \begin{tabular}{| l | l |}
    \hline
    \textbf{Date} & \textbf{Ranking}\\
	\hline
	April, 2009 & 5\\
	February, 2010 & 5\\
	March, 2010 & 4\\
	July, 2011 & 4\\
    August, 2011 & 4\\
    September, 2011 &4\\
    May, 2013 & 9\\
    August, 2013 & 10\\
    May, 2014 & 10\\
    June, 2014 & 10\\
    July, 2014 & 11\\
    \hline
    \end{tabular}
    \caption{Alexa Ranking for YouTube in Pakistan}
    \label{tab:alexapakistan}
    \vspace{-5pt}
\end{table}

\section{Results}\label{sec:results}
This section presents the results of our analysis. We first analyze the ranking
of YouTube (\S\ref{subsec:alexa}), followed by the popularity of individual
topics (\S\ref{subsec:youtube}). We then drill down into individual pieces of
restricted content (\S\ref{subsec:content}) and also examine the popularity of
anti-censorship tools (\S\ref{subsec:circumvention}).

\subsection{Website Ranking}\label{subsec:alexa}

YouTube has primarily borne the brunt of censorship in Pakistan. Most notably,
it was first briefly banned in 2008 and since September, 2012 has experienced an
indefinite blockage~\cite{Nabi:2013:TAO}. To gauge whether this censorship has
had any effect on the usage of the website in the country (or more specifically
its ranking), we make use of data from Alexa Web Rankings.
Unfortunately, Alexa no longer allows the user to query historical data for a
specific country and only displays the current ranking on its website. To work
around this, we use the \emph{Wayback
Machine}\footnote{\url{https://archive.org/web/}} from Internet Archive--which
caches previous versions of websites--to access previous rankings.
Table~\ref{tab:alexapakistan} presents the ranking of YouTube in Pakistan for
all Alexa snapshots from the Wayback Machine. It is clear that the ban on
YouTube in Pakistan has not affected its rankings or indirectly its usage in the
country. Specifically, in spite of the indefinite ban since 2012, the website
has been consistently amongst the top 11 websites in the country.

Similar to Pakistan, Turkey has also blocked YouTube multiple times in the last
few years. The longest ban lasted 30 months, from May, 2008 till October, 2010.
It also remained blocked between March, 2014 and June, 2014. In spite of this,
Alexa Rankings data in Table~\ref{tab:alexaturkey} shows that YouTube remained
amongst the top 10 most visited websites in Turkey during both censorship
periods.

\textbf{Limitations:} While the data we collected clearly shows a censorship
ineffectiveness trend, the wide gaps in the data do not permit us to claim that
the behaviour holds throughout. Nonetheless, it shows that at least for certain
months users were not deterred from accessing YouTube during the various
blackout periods in both countries. To examine whether there is a correlation
between content popularity and censorship, we next analyze data for specific
events.

\begin{table}[t]
\centering
    \begin{tabular}{| l | l |}
    \hline
    \textbf{Date} & \textbf{Ranking}\\
	\hline
	April, 2009 & 8\\
	June, 2009 & 8\\
    February, 2010 & 5\\
    September, 2011 & 3\\
    October, 2011 & 3\\
    April, 2012 & 3\\
    May, 2012 & 3\\
    May, 2013 & 3\\
    November, 2013 & 3\\
    March, 2014 & 4\\
    June, 2014 & 4\\
    July, 2014 & 4\\
    \hline
    \end{tabular}
    \caption{Alexa Ranking for YouTube in Turkey}
    \label{tab:alexaturkey}
    \vspace{-5pt}
\end{table}

\subsection{Topic Trends}\label{subsec:youtube}
\begin{figure}[t]
\centering
  \includegraphics[width=1\linewidth]{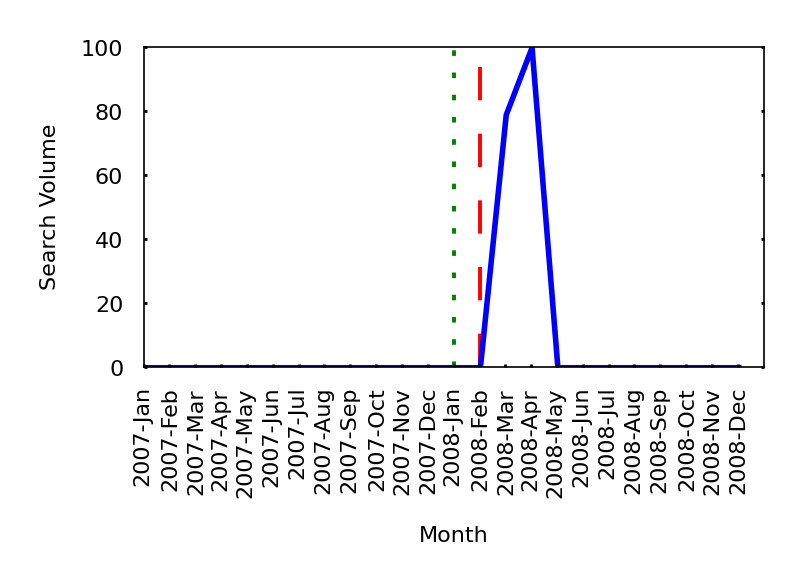}
  \caption{Search Volume for ``Fitna'' in Pakistan. The green dotted line marks
  the upload date on YouTube and the red dashed line marks its censorship.}
  \label{fig:fitna}
\end{figure}

Having shown that censorship has not affected the ranking of YouTube in Pakistan
and Turkey, we now dissect individual censorship events to study the
manifestation of the Streisand Effect. To this end, we use data from Google
Trends as a metric to gauge the popularity of restricted content. We first
examine data from Pakistan followed by Turkey.
As listed in Table~\ref{tab:pkevents}, YouTube was blocked in Pakistan in
February 2008 (\emph{PK1}) for a few days in retaliation to the controversial
movie ``Fitna''; the trailer for which was uploaded to YouTube in January of the
same year. In addition to Pakistan, several other countries, including
Indonesia, also blocked YouTube in response. Somewhat amusingly, Pakistan
restricted access to the website via BGP misconfiguration. This made the website
globally inaccessible for a large part of the Internet for nearly 2
hours~\cite{Nabi:2013:TAO}. Figure~\ref{fig:fitna} shows the Google search
volume for the term ``Fitna'' for \emph{users within Pakistan} in 2007 and 2008.
The dotted green line represents the month in which the video was uploaded while
the red dashed line marks the month when YouTube was blocked in Pakistan. The
blue line represents the actual search volume data. The data shows that the
popularity of the video spiked after it was restricted in Pakistan.

\begin{figure}[t]
\centering
  \includegraphics[width=1\linewidth]{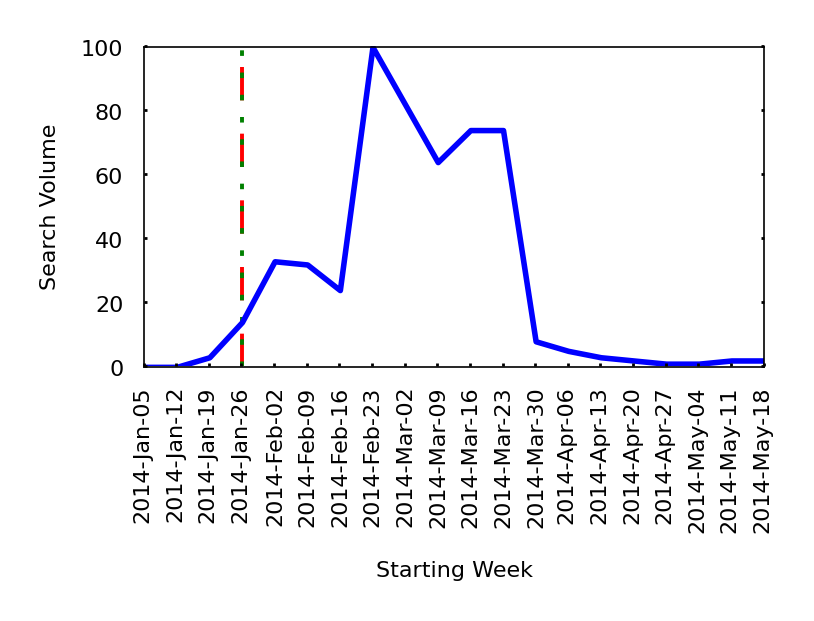}
  \caption{Search Volume for ``Haramzadeler'' in Turkey. Audio
  uploaded on SoundCloud: green dotted line, censorship: red dashed line.}
  \label{fig:haramzadeler}
\end{figure}

We also see a similar trend in Turkey. Since December 2013 a Twitter user with
ID ``Haramzadeler'' has been using the website to share audio clips and
documents implicating the ruling Turkish government in massive corruption
scandals. The collection also includes recordings of the Turkish Prime Minister
and his friends and family. In reaction to these exposŽs, the Turkish government
blocked access to SoundCloud, YouTube, and Twitter. In particular, SoundCloud
was blocked in January 2014 (\emph{TR2} in Table~\ref{tab:trevents}).
Figure~\ref{fig:haramzadeler} shows the Google search volume for
``Haramzadeler'' in Turkey. The results show that searches for the term peaked
as a result of the blockage of SoundCloud in January even though Haramzadeler
started uploading content months before that.

\begin{figure}[t]
\centering
  \includegraphics[width=1\linewidth]{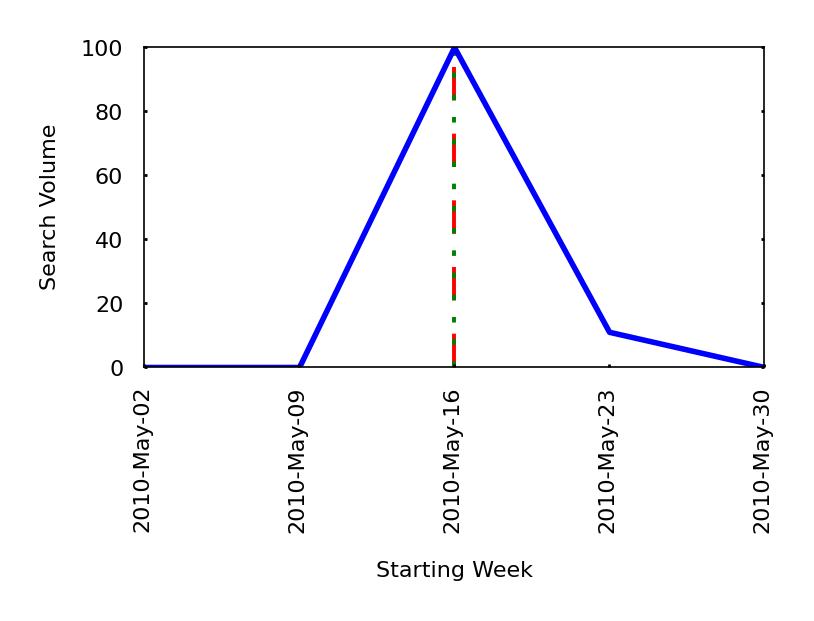}
  \caption{Search Volume for ``Everybody Draw Muhammad Day'' in Pakistan. Video
  upload: green dotted line, censorship: red dashed line.}
  \label{fig:draw}
\end{figure}

\begin{figure}[t]
\centering
  \includegraphics[width=1\linewidth]{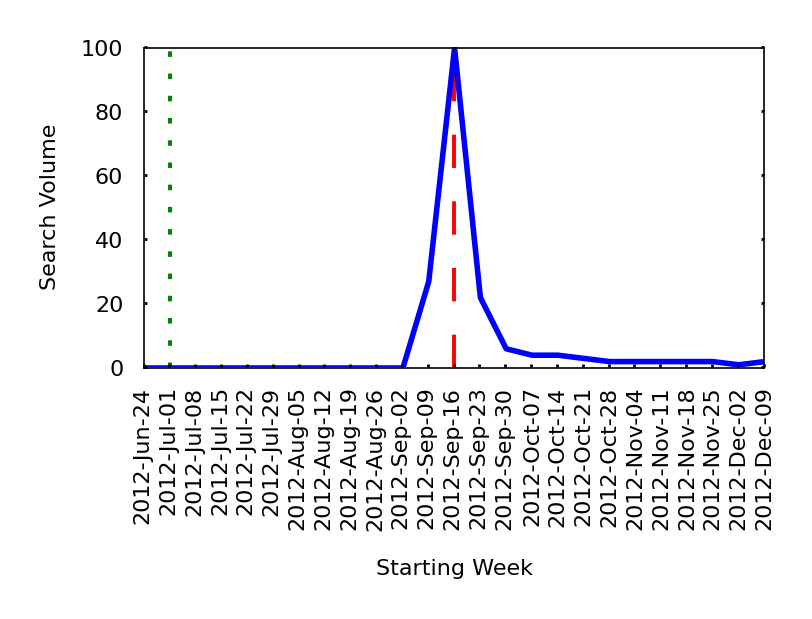}
  \caption{Search Volume for ``Innocence of Muslims'' in Pakistan. Video
  upload: green dotted line, censorship: red dashed line.}
  \label{fig:iom}
\end{figure}

\textbf{Limitations:} From the two examples above, it is fairly straight-forward
to conclude that there is a strong correlation between censorship and the
popularity of the censored content. It also suggests that this is not merely a
case of the classic ``correlation implies causation'' fallacy. Nonetheless, in
other cases this is harder to claim. For instance,
Figure~\ref{fig:draw}\footnote{Note that the x-axis now has fine-grained weekly
dated, unlike the monthly data in Figure~\ref{fig:fitna}.} shows Google Trends
data for Pakistan from May 2010 for ``Everybody Draw Muhammad Day'' (\emph{PK2}
in Table~\ref{tab:pkevents}). This event resulted in the blockage of YouTube for
a year. The finest granularity at which Google Trends data is available is on a
weekly scale, therefore it is hard to tell if the content went viral after
YouTube was blocked due to the fact that the content upload and censorship took
place during the same week and hence the overlap between the green and red
vertical lines. We see a similar trend in 2012, when YouTube was banned again
due to another controversial movie ``Innocence of Muslims'' (\emph{PK3} in
Table~\ref{tab:pkevents}). As before, the green dotted line marks the time when
the video was uploaded, the red dashed line marks the point at which the
censorship was enforced, and the blue line shows the search volume from within
Pakistan in Figure~\ref{fig:iom}. Again, from this data it is hard to conclude
if the censorship caused the popularity or vice versa.

\subsection{Affected Content}\label{subsec:content}
\begin{figure}[t]
\centering
  \includegraphics[width=1\linewidth]{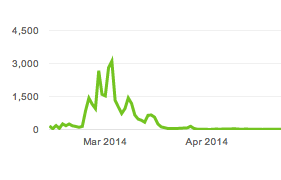}
  \vspace{-20pt}
  \caption{Daily video viewership statistics for an alleged conversation between
  the Turkish Prime Minister and a business tycoon. The x-axis is time and the
  y-axis is daily number of views. Graph was directly copied from YouTube.}
  \label{fig:erdogan}
\end{figure}

So far we have been able to show that, a) censorship has not affected website
rankings in Pakistan and Turkey, and b) topics spiked in popularity after they
were censored. The latter only applies to Google search popularity of the
content, not the content itself. Therefore, we now analyze YouTube statistics to
gauge whether we can pin-point the Streisand Effect at play for the actual
content. This task is complicated by two factors: a) YouTube only serves graphs
for statistics, not the underlying data, and b) most of the contentious videos
have now been taken down. As a result, we only focus on one particular clip
which was uploaded to YouTube in February 2014, wherein the Turkish Prime
Minister allegedly discusses construction permits for an area earmarked for a
forest with his friend who is a business tycoon and a real estate developer.
This video is amongst the set of content which resulted in the censorship of
both YouTube and Twitter (\emph{TR3} and \emph{TR4} in Table~\ref{tab:trevents})
in Turkey in March 2014. Figure~\ref{fig:erdogan} plots the daily views of the
said video. It is clear from the graph that even though the video was uploaded
in February, its popularity spiked in March, after YouTube was censored.

\textbf{Limitations:} While the example under consideration looks promising, it
is the only one for which we were able to obtain data. Therefore, a single video
is insufficient to extrapolate to other events.

\subsection{Circumvention}\label{subsec:circumvention}
\begin{figure*}[t]
\centering
  \includegraphics[width=1\linewidth]{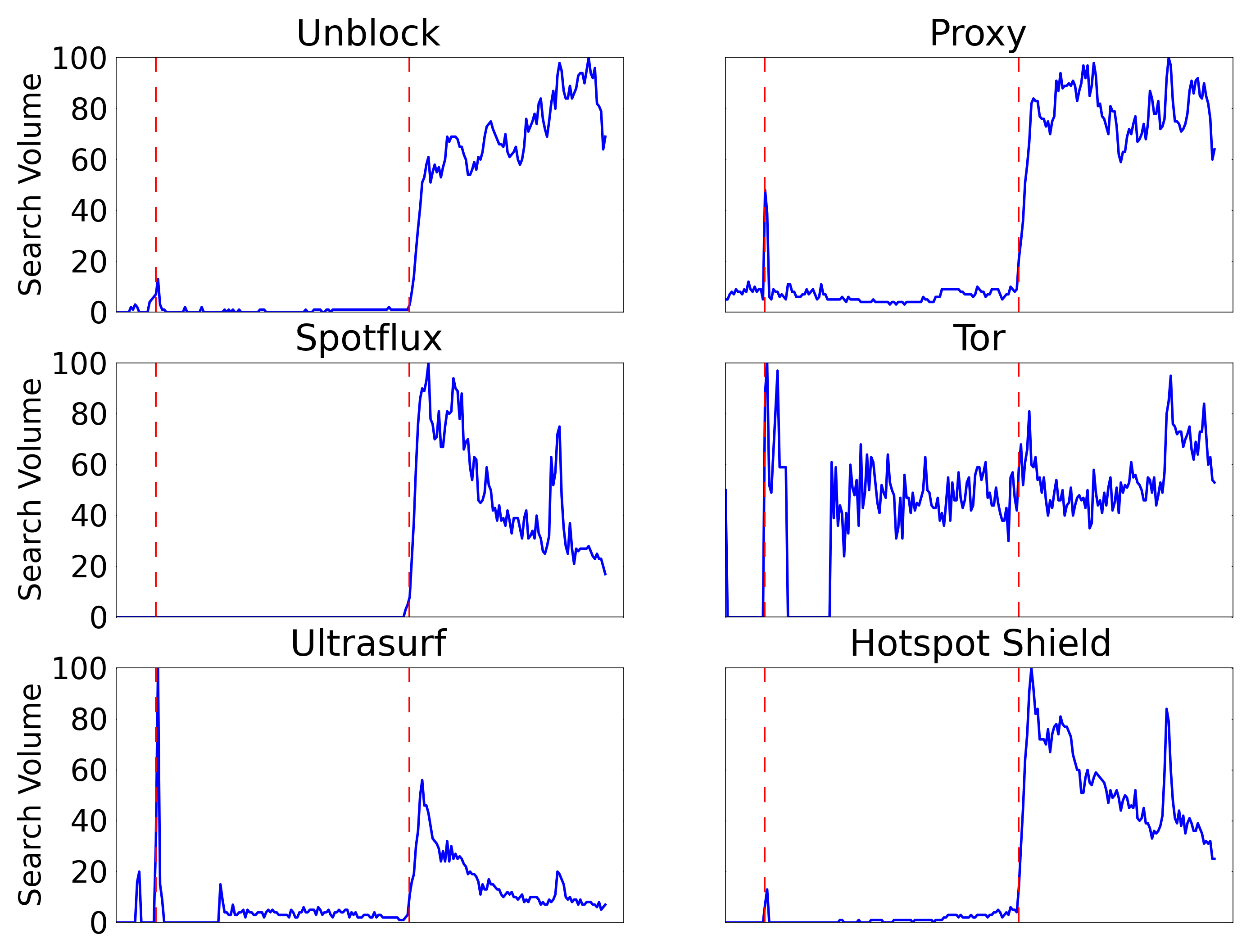}
  \caption{Search Volume for anti-censorship tools in Pakistan. The red lines
  mark \emph{PK2} and \emph{PK3}, respectively.}
  \label{fig:circumventpk}
\end{figure*}

\begin{figure*}[t]
\centering
  \includegraphics[width=1\linewidth]{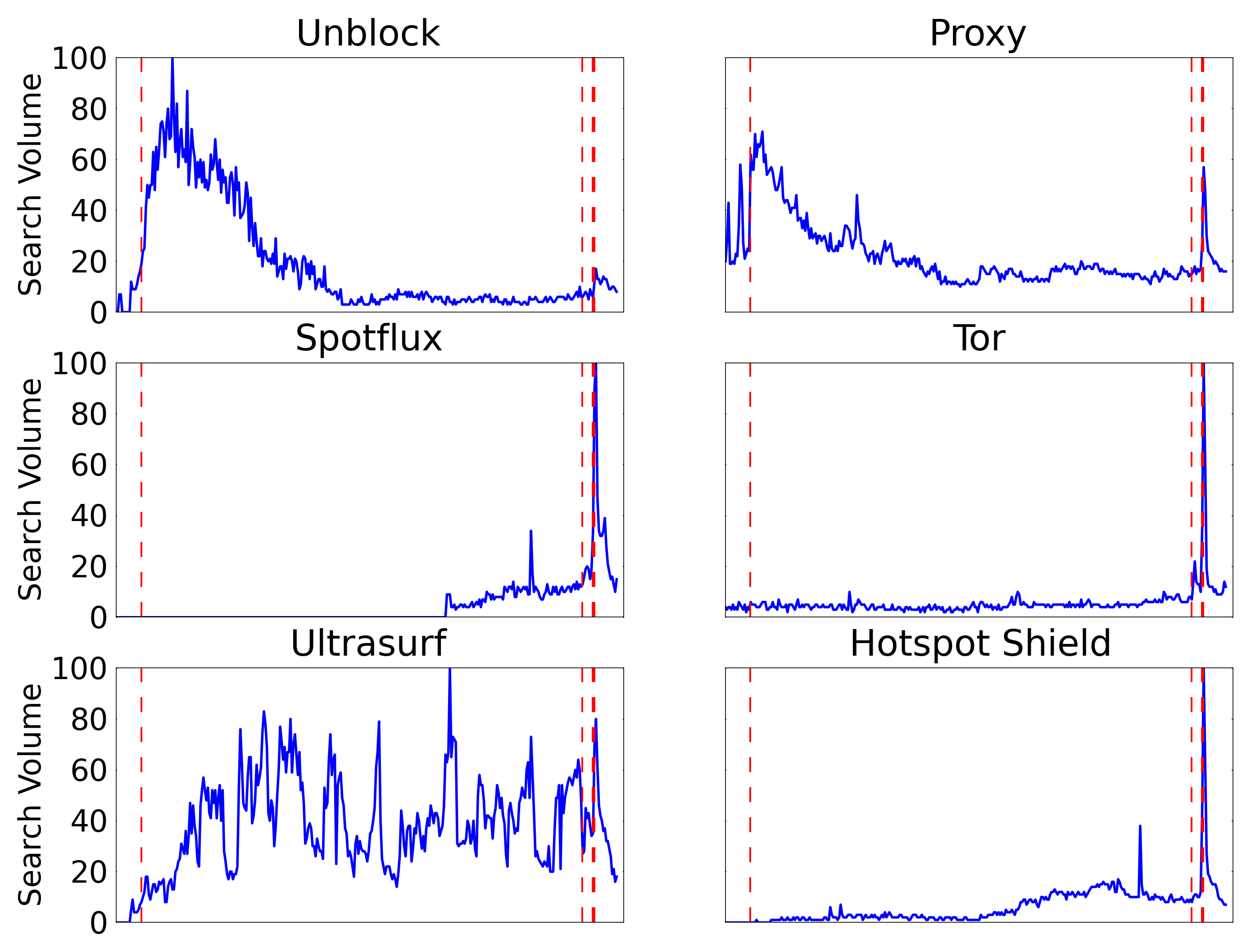}
  \caption{Search Volume for anti-censorship tools in Turkey. The red lines
  mark \emph{TR1} \emph{TR2}, \emph{TR3}, and \emph{TR4}, respectively.}
  \label{fig:circumventtr}
\end{figure*}

Our analysis thus far shows that censorship has not deterred users from
accessing restricted content. In fact, in certain cases the censorship has had
the side-effect of popularising the content. Bearing this in mind, the next
natural question is to ask how users are accessing the restricted content. In
case of Pakistan, in prior work~\cite{Nabi:2013:TAO} we learned that users were
actively making use of anti-censorship tools, such as VPNs and Tor to circumvent
censorship. To determine whether there is a correlation between censorship and
the use of anti-censorship tools, we again rely on data from Google Trends.
Specifically, we note the search volume for ``Spotflux'', ``Tor'',
``Ultrasurf'', and ``Hotspot Shield''. We also add two generic circumvention
terms viz ``Unblock'' and ``Proxy''. The data is visualized in
Figure~\ref{fig:circumventpk}, the two red dashed lines mark \emph{PK2} and
\emph{PK3}, respectively. We see that the popularity of these tools spiked as a
result of the two censorship events. Note that Spotflux was not available before
\emph{PK2}, that is why there is no spike in its search volume. We also
conducted a similar study for Turkey, with the results presented in
Figure~\ref{fig:circumventtr}. In this case, the four red lines represent
\emph{TR1}, \emph{TR2}, \emph{TR3}, and \emph{TR4}, respectively. Again, we see
that the popularity of these tools is positively correlated with the four
censorship events. Unlike Pakistan, where Tor is popular in the entire time
series, in Turkey its usage seems to have started only after \emph{TR3} and
\emph{TR4}.

\section{Discussion}\label{sec:discussion}
Using limited and sparse data from multiple sources we have been able to show
that not only does censorship not work but it also inadvertently causes
restricted content to become popular. This is a promising line of work but
before Internet activists around the world can claim a convincing victory
against the spectre of censorship, we need to analyze more comprehensive data.
Towards this end, we now discuss how additional data might be useful and we also
call upon organizations, such as Alexa and Google, who might hold this data to
share it with the research community so that they can comb through it. 

\subsection{Open Data Against Censorship}
In \S\ref{subsec:alexa}, using data from Alexa Web Rankings, we argued that
website level censorship has not diminished the usage of YouTube in Pakistan and Turkey
for certain months. But the lack of monthly data throughout the events deters us
from claiming the consistency of the trend. Similarity, in
\S\ref{subsec:youtube}, using Google Trends data, we showed that certain pieces
of content went viral \emph{after} being censored, arguing for a cause and
effect relationship between the latter and the former. While this trend was
clear for \emph{PK1} and \emph{TR2}, the weekly granularity of the data prevents
us from verifying this for \emph{PK2} and \emph{PK3}. Fine-grained daily data
would be sufficient for teasing out the presence of the Streisand Effect. In the
same vein, data about the usage of other videos on YouTube would be effective in
inspecting the effect for individual pieces of content. In addition, data from
ISPs and Content Distribution Networks would be useful in examining access
patterns during periods of censorship at a finer granularity. We hope that all
of these organizations that are sitting on top of these silos of data will make
it publicly available, similar to the filtering
data\footnote{\url{https://opennet.net/research/data}} from the OpenNet
Initiative.

\subsection{The Streisand Effect as a means to an end}\label{subsec:means} 
The Streisand Effect in a large number of instances showcases the futility of
online censorship. Primarily, it shows that banning content on the Internet generates
more interest in the content and increases its circulation. Cognizance of this
phenomenon is slowly driving public policy and discourse. A recent example of
this is the aftermath of the May, 2014 European Court of Justice ruling in the
Google vs Gonz\'{a}lez case. Under this ``right to be forgotten'' ruling,
individuals have the right to request search engines to remove personal data
from search results~\cite{EC:2014:FOT}. In the ensuing debate, EU countries
have been weighing the implications and practicality of the EU-wide ``General
Data Protection Regulation''. Italy for instance in one of its proposals assumed
that individuals want an umbrella removal of their personal information, which
potentally requires a search engine such as Google to reach out to third
parties, e.g. individual social network users or people with personal webpages,
to cascade information deletion. In reaction, Poland opined that the assumption
is ill-founded and the required multimodal removal effort will most likely
trigger the Streisand Effect as it will draw more attention to the
content~\cite{CEU:2014:PFT}. Under Poland's proposition, users should be able
to choose which information to remove and from where. For example, a user might
want a certain piece of information to be removed from Google search results but
still be available on a blog. Therefore, awareness of the effect will be
influential in defining digital rights policy in the EU. In a similar vein, we
believe that governments that directly censor information on the Internet need
to be mindful of inadvertent consequences of their actions. For example,
countries like Pakistan and Turkey, regularly ban online content to safeguard
the religious and social sensibilities of their citizens. Our examination of
this category of censorship in this paper has shown that it has largely been
ineffective. Knowledge of this observation should prompt state-level censors to
rethink their position and explore alternative mechanisms to deal with such
issues--for example, by directly engaging with their citizens and considering
their points of view.

While the Streisand Effect is a handy instrument to keep censorship in check, it
is only one of the many means to an end, not an end in itself. The end being an
open, universally accessible Internet. To begin with, the effect does not
manifest in all instances of censorship. The rampant presence of censorship
around the world is a testament to this~\cite{Burnett:2013:MSI}. Secondly, the
goal of this paper is not to argue that online activists need to become
complacent and actually embrace online censorship as it, due to the Streisand
Effect, achieves the goal of proving its own ineffectiveness. On the contrary,
anti-censorship efforts need to be beefed up in terms of political activism and
campaigning, design of appropriate tools, and identification and measurement of
Internet restrictions to keep up with the increase in country-level
censorship~\cite{Bicchierai:2013:GGC}. Specifically, once content has been
blocked, anti-censorship tools such as VPNs and proxies need to be employed.
Without these tools the category of the Streisand Effect presented in this paper
would not originate in the first place. Furthermore, the initial study and
measurement of Internet censorship enables the design of tools to counteract it.
For instance, ignoring TCP RST packets as a method to bypass the Great Firewall
of China came about as a result of studying its modus
operandi~\cite{Clayton:2006:IGF}.
Finally, the efforts of Internet freedom activists in recent years have been
effective in raising awareness. In some cases--for instance, in
Turkey~\cite{Reuters:2014:TLT}--they have even succeeded in getting content
unblocked through legal recourses. Overall, the Streisand Effect would not exist
if all of the above efforts were abandoned. In essence, the aim of this paper is
to raise awareness about the Streisand Effect in the context of state-level
censorship rather than to argue for it as an all-encompassing anti-censorship
mechanism.

\section{Conclusion and Future Work}\label{sec:conclusion}
By making use of data from Google Trends, YouTube Statistics, and Alexa Web
Rankings, we were able to not only show that censorship is ineffective but
rather that it also has the inadvertent effect of making restricted content go
viral. Specifically, we show that the national ranking of YouTube in Pakistan
and Turkey has remained largely the same in spite of multiple censorship events.
In addition, Google Trends and YouTube Statistics numbers show that blocked
videos spiked in popularity after they were censored.
This paper was just an initial window into the interplay of the Streisand Effect
and state-level censorship. In the future, we aim to extend our study to other
countries. In addition, we hope to make use of other datasets to augment our
current findings. Finally, Google Trends data also breaks down search volume by
region. It would interesting to work out say if censorship circumvention is an
urban or rural phenomenon in Pakistan and Turkey.

\section*{About the Author}
Zubair Nabi is a Research Staff Member in the Big Data Systems and Analytics
Group at IBM Research, Ireland. His interests include data intensive computing,
technology for the developing world, and Internet accessibility. His work has
been featured in MIT Technology Review, CNET, Asian Scientist, and SciDev.net,
among others. Zubair has an MPhil in Computer Science from the University of
Cambridge.


\balance
{\footnotesize \bibliographystyle{acm}
\bibliography{foci}}

\end{document}